# Impossibility of a perpetuum mobile based on the nano-confinement effect on chemical equilibrium

Leonid Rubinovich

Department of Chemistry, Ben-Gurion University

Beer-Sheva, ISRAEL

Subsequent to recent prediction of the nano-confinement effect on chemical equilibrium (*1*), a case of a quasi-closed nanosystem is considered in the present work. Here the term "quasi-closed" means that molecule exchange with the environment is so weak that the nanosystem has enough time to approach the equilibrium.

Because of the nano-confinement induced difference in the equilibrium constants of the nanosystem and the environment, compositions of molecular flows entering and going out of the nanosystem seem to be different (Fig.1). Can these perpetual flows move a turbine having blades with different penetrability for different types of molecules? If the answer would be yes, a perpetuum mobile could be constructed.

Another perpetuum mobile project is shown schematically in Fig.2. Because of the nano-confinement induced increase of the equilibrium constant in the nanosystem, the exothermic reaction in pores goes closer to completion (to the products). Can the produced heat be used to generate electric current?

Since these devices contradict the second law of thermodynamics, some explanations are needed for the fact that they will not work. Fluctuations in the number of molecules and in the composition of the nanosystem seem to be the reason. For example, in case of extremely small system comprising maximum two molecules, reaction is impossible if the number of molecules is zero, one, or if pairs (HD, $H_2$) or (HD, $D_2$) present. This can diminish the flow of the product out of the nanosystem and make it equal to the flow to the nanosystem.

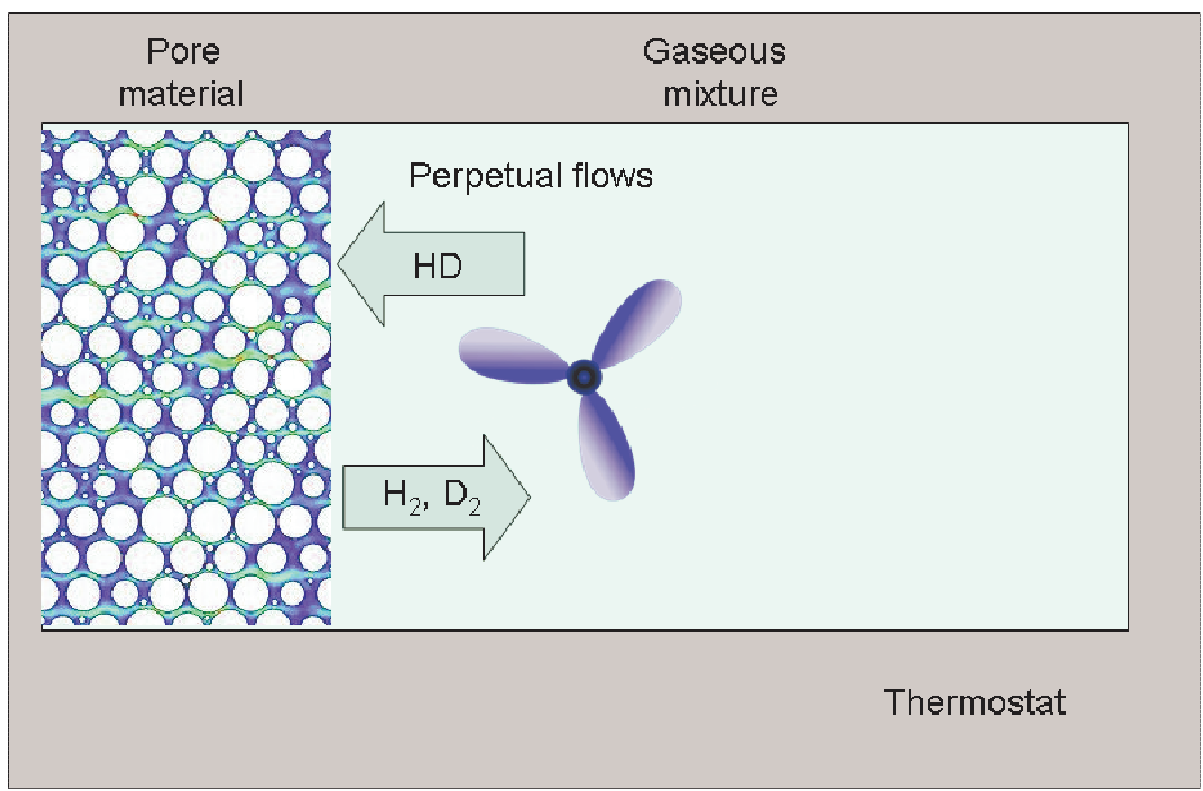

Fig.1. Can perpetual flows move a turbine having blades with different penetrability for different molecules?

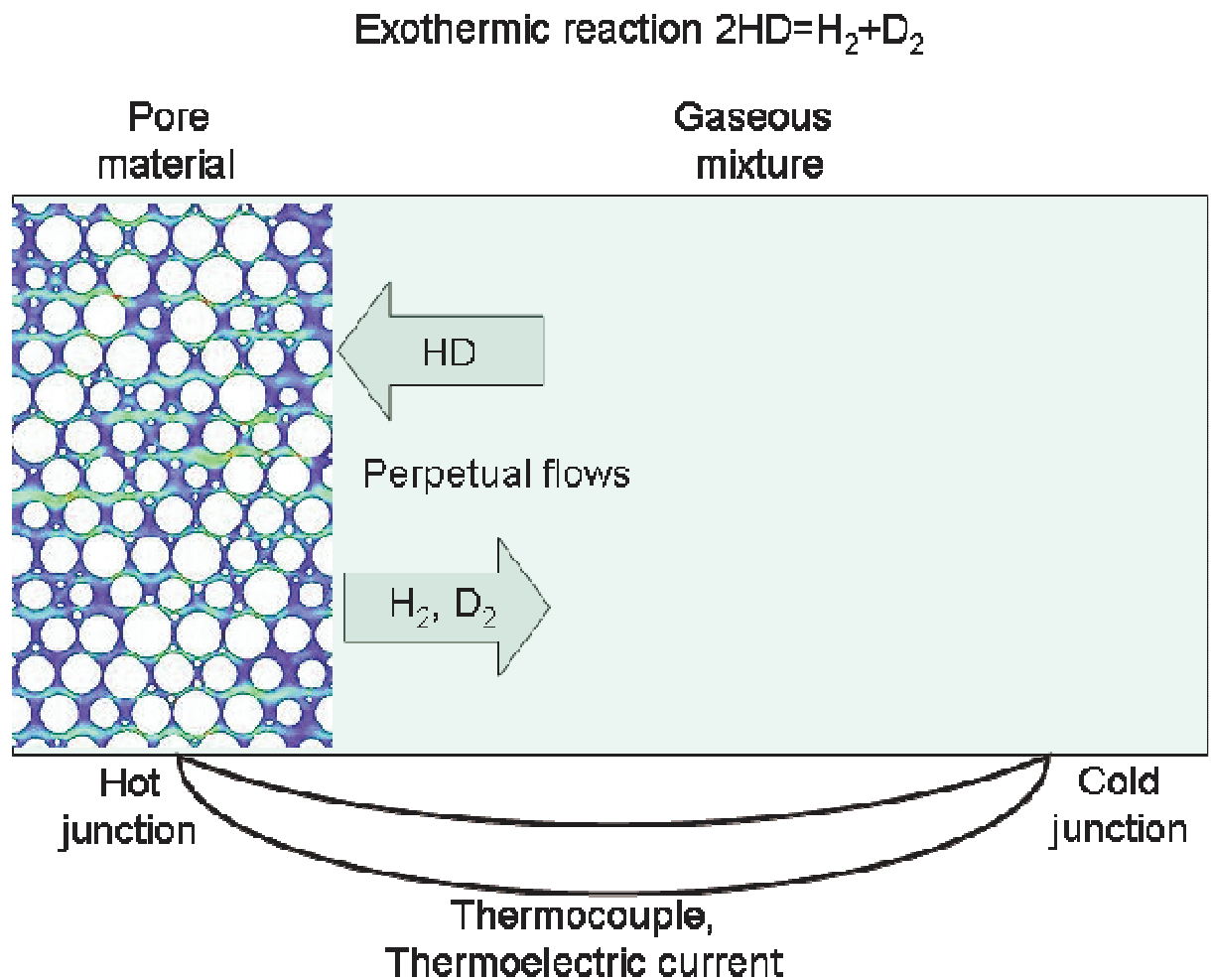

Fig.2. Can the smallness effect be used to generate electric current?

1. M. Polak, L. Rubinovich, *Nano Letters* **8**, 3543 (Oct, 2008).